\documentclass[12pt]{article}
\usepackage{epsfig}
\usepackage{multirow}
\usepackage{cite}

\newcommand{\mysection}{\setcounter{equation}{0}\section}

\def\beq{\begin{equation}}
\def\eeq{\end{equation}}
\def\beqa{\begin{eqnarray}}
\def\eeqa{\end{eqnarray}}
 
\newlength{\dinwidth} \newlength{\dinmargin}
\begin{document}

\begin{center}
{\Large \bf Three-loop cusp anomalous dimension and a conjecture for $n$ loops}
\end{center}
\vspace{2mm}
\begin{center}
{\large Nikolaos Kidonakis}\\
\vspace{2mm}
{\it Department of Physics, Kennesaw State University,\\
Kennesaw, GA 30144, USA}
\end{center}
 
\begin{abstract}
I present analytical expressions for the cusp anomalous dimension in QCD through three loops in terms of elementary functions and ordinary polylogarithms. I observe interesting relations between the results at different loops and provide a conjecture for the $n$-loop cusp anomalous dimension in terms of the lower-loop results. I also present numerical results and simple approximate formulas for the cusp anomalous dimension relevant to top-quark production.
\end{abstract}

\mysection{Introduction}

The cusp anomalous dimension is a fundamental object in quantum field theory, and QCD in particular, that controls the infrared behavior of perturbative scattering amplitudes \cite{AMP,BNS,IKR,KR,NK2loop,NK2lc,CHMS,HH,GHKM}.  Its study and related techniques have been useful in a large variety of subjects in perturbative QCD, including soft anomalous dimensions and infrared structure in hard-scattering processes (see e.g. \cite{NKGS,KOS,ADS,BN,GM,LD,MSS,BFS,FNPY,DGM,GLSW,NKst,NKtt,GSW,EG,FGHMW,ADG} and references therein). 

The first two-loop calculation of the cusp anomalous dimension was performed in \cite{KR} and the result included a few uncalculated integrals. An independent calculation, specifically targeted towards heavy-quark production, appeared later in \cite{NK2loop} (see also \cite{NK2lc}) and provided an explicit result in terms of elementary functions, dilogarithms, and trilogarithms. 

Recently, the three-loop result for the cusp anomalous dimension in QCD was calculated and presented in \cite{GHKM}. The expression is much more complicated and is given in terms of a large number of harmonic polylogarithms, each of which is defined iteratively and involves multiple integrals with up to five integrations. In this paper we use the results of \cite{GHKM} and present the cusp anomalous dimension in a different but fairly compact expression involving ordinary polylogarithms. All but a few of the harmonic polylogarithms can be expressed in terms of elementary functions and ordinary polylogarithms, with the remaining few involving single integrals (complete results for those calculations are given in the Appendix). 
 
We find that the structure of the results is more transparent in the new expressions. In fact our expressions point to relations among the cusp anomalous dimensions at different number of loops and suggest a pattern. Thus, a conjecture is made that expresses the $n$-loop result in terms of results through $n-1$ loops, and we use the conjecture to provide some predictions for the four-loop and five-loop cusp anomalous dimensions in terms of known and some unknown functions.
 
The cusp anomalous dimension is a basic ingredient for calculations of soft anomalous dimensions for various processes, including top-quark production. Numerical results are shown, and simple but excellent approximations are also derived for the cusp anomalous dimension through three loops. We hope that the explicit analytical and numerical results presented in this paper will be useful in higher-loop calculations of soft anomalous dimensions. For example, top-quark production would be one important application.

In the next section we provide explicit results for the cusp anomalous dimension through three loops. In Section 3 we present numerical results which are particularly relevant to top-quark production, and we construct simple approximations to the full analytical result. Section 4 presents the conjecture for $n$ loops. We conclude in Section 5. Details and results for the harmonic polylogarithms and other functions are provided in the Appendix.

\mysection{Cusp anomalous dimension at three loops}

\begin{figure}
\begin{center}
\includegraphics[width=10cm]{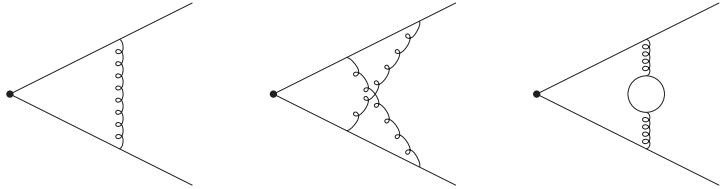}
\caption{Typical one-loop (left) and two-loop (middle and right) diagrams for the cusp anomalous dimension.}
\label{loop12plot}
\end{center}
\end{figure}

\begin{figure}
\begin{center}
\includegraphics[width=10cm]{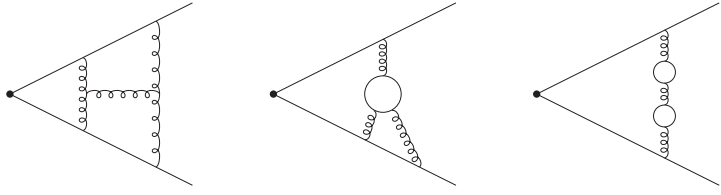}
\caption{Typical three-loop diagrams for the cusp anomalous dimension.}
\label{loop3plot}
\end{center}
\end{figure}

The perturbative expansion of the cusp anomalous dimension in QCD is written as 
\beq
\Gamma_{\rm cusp}=\sum_{n=1}^{\infty} \left(\frac{\alpha_s}{\pi}\right)^n \Gamma^{(n)}
\eeq 
where $\alpha_s$ is the strong coupling.
Some typical diagrams contributing at one and two loops are shown in Fig. \ref{loop12plot}, and at three loops in Fig. \ref{loop3plot}, with the eikonal (Wilson) lines representing the massive quarks. The cusp angle $\theta$ is given by $\theta=\cosh^{-1}(v_i\cdot v_j/\sqrt{v_i^2 v_j^2})$ where $v_i^{\mu}$ and $v_j^{\mu}$ are velocity vectors for quarks $i$ and $j$. In dimensional regularization with 
$4-\epsilon$ dimensions, the cusp anomalous dimension is the coefficient of the ultraviolet $1/\epsilon$ pole arising from the eikonal diagrams. See Ref. \cite{NK2loop} for more details. 

The one-loop expression for the cusp anomalous dimension is given in terms of the cusp angle $\theta$ by
\beq
\Gamma^{(1)}=C_F (\theta \coth\theta-1)
\label{1loop}
\eeq  
where $C_F=(N_c^2-1)/(2N_c)$ with $N_c$ the number of colors.

The two-loop expression for the cusp anomalous dimension can be written as  
\beqa
\Gamma^{(2)}&=&\frac{K}{2} \, \Gamma^{(1)}
+\frac{1}{2}C_F C_A \left\{1+\zeta_2+\theta^2 
-\coth\theta\left[\zeta_2\theta+\theta^2
+\frac{\theta^3}{3}+{\rm Li}_2\left(1-e^{-2\theta}\right)\right] \right. 
\nonumber \\ && \hspace{30mm} \left.
{}+\coth^2\theta\left[-\zeta_3+\zeta_2\theta+\frac{\theta^3}{3}
+\theta \, {\rm Li}_2\left(e^{-2\theta}\right)
+{\rm Li}_3\left(e^{-2\theta}\right)\right] \right\}
\label{2loop}
\eeqa
where $K=C_A (67/18-\zeta_2)-10 T_F n_f/9$ with $C_A=N_c$, $T_F=1/2$, and $n_f$ the number of light quark flavors. The expression shown in Eq. (\ref{2loop}) is even simpler than the one presented in Ref. \cite{NK2loop}. $\Gamma^{(2)}$ involves a few elementary functions, $\zeta_2$ and $\zeta_3$ constants, two dilogarithms, and a trilogarithm (see the Appendix for definitions of the zeta constants and ordinary polylogarithms). The $n_f$ terms in $\Gamma^{(2)}$ arise from quark loops, e.g. as shown in the rightmost diagram of Fig. \ref{loop12plot}.

The three-loop result has been presented in \cite{GHKM} in terms of a large number of harmonic polylogarithms of weight up to 5, each one of which involves up to quintiple integrals. After explicit evaluation of those integrals, as detailed in the Appendix, and after some rearrangements and grouping of terms, the three-loop cusp anomalous dimension can be rewritten compactly as 
\beqa
\Gamma^{(3)}&=& C_F C_A^2 c_1-\frac{\Gamma^{(1)}}{27} T_F^2 n_f^2
\nonumber \\ &&
{}+\left\{-\frac{10}{9} \Gamma^{(2)}+\frac{\Gamma^{(1)}}{9}
\left[5K+C_F\left(9 \zeta_3-\frac{165}{16}\right)
+C_A\left(5 \zeta_2-\frac{21}{2} \zeta_3-\frac{209}{24}\right)\right]\right\} T_F n_f
\nonumber \\
\label{3loop}
\eeqa
where the $n_f^2$ terms are proportional to $\Gamma^{(1)}$, and the $n_f$ terms have been written compactly in terms of $\Gamma^{(1)}$ and $\Gamma^{(2)}$.
Note that the $n_f$ and $n_f^2$ terms are given by diagrams such as the middle and right graphs, respectively, of Fig. \ref{loop3plot}. The term $c_1$ is independent of $n_f$ and it is given by  
\beqa
c_1&=& \quad
-\frac{199}{288}+\frac{67}{18}\zeta_2-\frac{7}{12}\zeta_3
-\frac{15}{4}\zeta_4
-\frac{\zeta_2}{2}\theta+\left(\frac{29}{18}-\frac{\zeta_2}{2}\right)\theta^2
-\frac{\theta^3}{12}-\frac{\theta^4}{24}
\nonumber \\ && \quad 
+\frac{\theta^2}{4} \ln \left(1-e^{-2 \theta}\right)
-\frac{3}{4} \theta {\rm Li}_2\left(e^{-2 \theta}\right)
-\frac{5}{8} {\rm Li}_3\left(e^{-2 \theta}\right)
\nonumber \\ && \hspace{-10mm}
+\coth\theta \left\{-\frac{\zeta_3}{4}+\frac{15}{8}\zeta_4
+\left(\frac{245}{96}-\frac{29}{9}\zeta_2-\frac{\zeta_3}{24}
+\frac{15}{4}\zeta_4\right) \theta +\left(-\frac{29}{18}+\frac{3}{2}\zeta_2\right)
\left(\theta^2+\frac{\theta^3}{3}\right) \right.
\nonumber \\ && \quad \quad
+\frac{7}{24}\theta^4 +\frac{\theta^5}{24}
+\frac{1}{2} \left(\theta^2+\theta^3\right) \ln\left(1-e^{-2 \theta}\right)
+\left(-\frac{29}{18}+\frac{\zeta_2}{2}\right) 
{\rm Li}_2\left(1-e^{-2 \theta}\right)
\nonumber \\ && \quad \quad \left.
-\frac{3}{4} \theta^2 {\rm Li}_2\left(e^{-2 \theta}\right)
+\frac{1}{4} (1-7\theta){\rm Li}_3\left(e^{-2 \theta}\right)
+\frac{1}{2} {\rm Li}_3\left(1-e^{-2 \theta}\right) 
-\frac{15}{8} {\rm Li}_4\left(e^{-2 \theta}\right) \right\}
\nonumber \\ && \hspace{-10mm}
+\coth^2\theta \left\{-\frac{67}{36} \zeta_3-\frac{19}{8}\zeta_4+\frac{3}{2}\zeta_5
+\left(\frac{67}{36}\zeta_2+\frac{3}{2} \zeta_3-\frac{25}{8} \zeta_4 \right) 
\theta+\left(\frac{\zeta_3}{4}-\zeta_2\right) \theta^2
\right.
\nonumber \\ && \quad \quad
+\left(\frac{67}{108}-\frac{5}{6}\zeta_2\right) \theta^3
-\frac{\theta^4}{4}-\frac{11}{120}\theta^5 
-\left(-\zeta_3+\zeta_2 \theta+\zeta_2 \theta^2
+\frac{\theta^3}{3}+\frac{\theta^4}{6}\right) \ln\left(1-e^{-2 \theta}\right)
\nonumber \\ && \quad \quad
-\theta^2 \ln^2\left(1-e^{-2 \theta}\right)
-\theta \ln^3\left(1-e^{-2 \theta}\right)
-\frac{1}{8} \ln^4\left(1-e^{-2 \theta}\right)
\nonumber \\ && \quad \quad
+\left[\frac{\zeta_2}{2}+\left(\frac{67}{36}+\frac{\zeta_2}{2}\right)\theta-2\theta^2+\frac{\theta^3}{12}-\theta \ln\left(1-e^{-2 \theta}\right) \right]
{\rm Li}_2\left(e^{-2 \theta}\right)
\nonumber \\ && \quad \quad
-\frac{1}{4} {\rm Li}_2^2\left(e^{-2 \theta}\right)
+\frac{1}{2} \ln^2\left(1-e^{-2 \theta}\right) {\rm Li}_2\left(1-e^{-2 \theta}\right)
+\frac{1}{4} {\rm Li}_2^2\left(1-e^{-2 \theta}\right)
\nonumber \\ && \quad \quad
-\frac{1}{2} \ln^2\left(e^{2 \theta}-1\right)
{\rm Li}_2\left(\frac{1}{1-e^{2 \theta}}\right) 
+\left[\frac{67}{36}-\frac{3}{2} \theta -\frac{\theta^2}{4}
-\ln\left(1-e^{-2 \theta}\right)\right]{\rm Li}_3\left(e^{-2 \theta}\right)
\nonumber \\ && \quad \quad
-\left[\theta+\ln\left(1-e^{-2 \theta}\right) \right] 
{\rm Li}_3\left(1-e^{-2 \theta}\right)
-\left[2 \theta+\ln \left(1-e^{-2 \theta}\right) \right] 
{\rm Li}_3\left(\frac{1}{1-e^{2 \theta}}\right)
\nonumber \\ && \quad \quad \left.
-\frac{9}{8} \theta {\rm Li}_4\left(e^{-2 \theta}\right)
+{\rm Li}_4\left(1-e^{-2 \theta}\right)-{\rm Li}_4\left(\frac{1}{1-e^{2 \theta}}
\right)-\frac{3}{2} {\rm Li}_5\left(e^{-2 \theta}\right) \right\}
\nonumber \\ && \hspace{-10mm}
+\frac{1}{4} \left[A(\theta)-A(0)+B(\theta)-B(0)\right] 
\label{c1}
\eeqa
where
\beqa
A(\theta)&=&
\coth^3\theta \left\{-3 \zeta_5
+4 \zeta_4 \theta-3 \zeta_3 \theta^2+\frac{4}{3} \zeta_2 \theta^3
+\frac{\theta^5}{5} \right.
\nonumber \\ && \quad \quad
+\frac{2}{3} \theta^3 {\rm Li}_2\left(e^{-2 \theta}\right) 
+\theta^2 {\rm Li}_3\left(e^{-2 \theta}\right) 
+2 \theta {\rm Li}_4\left(e^{-2 \theta}\right)
+3 {\rm Li}_5\left(e^{-2 \theta}\right) 
\nonumber \\ && \quad \quad \left.
+H_{1,1,0,0,1}(1-e^{-2\theta})+H_{1,0,1,0,1}(1-e^{-2\theta}) \right\}
\eeqa
and
\beqa
B(\theta)&=&
\frac{e^{\theta}}{e^{2 \theta}-1}
\left\{-2 \zeta_2 \zeta_3 + 2 \zeta_3 \theta^2 
+\left(\frac{3}{2} \zeta_4 -\frac{\theta^4}{6} \right)
\ln\left(e^{\theta}-1\right) \right.
\nonumber \\ && \quad \quad
+\left(-\frac{3}{2} \zeta_4-2\zeta_3 \theta +\frac{\theta^4}{6} \right)
\ln\left(e^{\theta}+1\right)  
+2 \zeta_3 \left[{\rm Li}_2\left(-e^{-\theta}\right)
+{\rm Li}_2\left(1-e^{-\theta}\right)\right]
\nonumber \\ && \quad \quad
+\frac{2}{3} \theta^3 \left[{\rm Li}_2\left(e^{-\theta}\right)
-{\rm Li}_2\left(-e^{-\theta}\right)\right]
+2 \theta^2 \left[{\rm Li}_3\left(e^{-\theta}\right)
-{\rm Li}_3\left(-e^{-\theta}\right)\right]
\nonumber \\ &&  \quad \quad 
+4 \theta \left[{\rm Li}_4\left(e^{-\theta}\right)
-{\rm Li}_4\left(-e^{-\theta}\right)\right]
+4 {\rm Li}_5\left(e^{-\theta}\right)
-4 {\rm Li}_5\left(-e^{-\theta}\right)
\nonumber \\ && \quad \quad \left. 
+4 \left[H_{1,0,1,0,0}(e^{-\theta})+H_{-1,0,1,0,0}(e^{-\theta})
-H_{1,0,-1,0,0}(e^{-\theta})-H_{-1,0,-1,0,0}(e^{-\theta}) \right]
\right\} \, .
\nonumber \\
\eeqa
We note that the functions $A(\theta)$ and $B(\theta)$ involve some weight 5 harmonic polylogarithms that cannot be expressed in terms of ordinary polylogarithms. However they can be reduced to single integrals of elementary functions and ordinary polylogarithms as shown in the Appendix. All other terms in $c_1$ involve elementary functions, $\zeta_2$, $\zeta_3$, $\zeta_4$, and $\zeta_5$ constants, as well as ordinary polylogarithms ${\rm Li}_k$ with $k=2$, 3, 4, 5. In particular, we note that $\coth\theta$ terms appear at one, two, and three loops; and $\coth^2\theta$ terms appear at two and three loops; and all these terms can be written in terms of elementary functions and standard polylogarithms. At three loops we also have $\coth^3\theta$ terms, in the function $A(\theta)$, as well as additional terms in the function $B(\theta)$ which are not expressible in that manner. If we try to express the weight 5 harmonic polylogarithms in $A$ and $B$ in terms of ordinary polylogarithms then we can reduce the result to single integrals, and the functions $A(\theta)$ and $B(\theta)$ can be written alternatively as
\beqa
A(\theta)&=&\coth^3\theta \left\{3 \zeta_5
+\frac{19}{2}\zeta_4 \theta-3 \zeta_3 \theta^2+\frac{4}{3} \zeta_2 \theta^3
+\frac{\theta^5}{5} +\frac{\theta}{2} \ln^4\left(1-e^{-2 \theta}\right) \right.
\nonumber \\ && \quad \quad
+\left(-2 \zeta_3+2\zeta_2\theta +\frac{2}{3} \theta^3\right)
{\rm Li}_2\left(e^{-2 \theta}\right)
-\ln^3\left(1-e^{-2 \theta}\right) {\rm Li}_2\left(1-e^{-2 \theta}\right)
\nonumber \\ && \quad \quad
+\left(2\zeta_2-3 \theta^2\right) {\rm Li}_3\left(e^{-2 \theta}\right)
+3 \ln^2\left(1-e^{-2 \theta}\right) {\rm Li}_3\left(1-e^{-2 \theta}\right)
-6 \theta {\rm Li}_4\left(e^{-2 \theta}\right)
\nonumber \\ && \quad \quad
-6 \ln\left(1-e^{-2 \theta}\right) {\rm Li}_4\left(1-e^{-2 \theta}\right)
-3  {\rm Li}_5\left(e^{-2 \theta}\right)
+6 {\rm Li}_5\left(1-e^{-2 \theta}\right)
\nonumber \\ && 
+\int_0^{1-e^{-2\theta}} \left[-\ln(1-z) \ln^3 z
+\frac{1}{2} \ln^2(1-z) \ln^2 z 
-\ln^2 z \, {\rm Li}_2(z)  \right. 
\nonumber \\ && \quad \quad 
-\ln(1-z) \ln z \, {\rm Li}_2(1-z) 
+\ln^2\left(\frac{1-z}{z}\right) {\rm Li}_2\left(\frac{z-1}{z}\right)
\nonumber \\ && \quad \quad 
-\frac{1}{2}{\rm Li}_2^2(z)+\frac{1}{2}{\rm Li}_2^2(1-z)
-\ln(1-z){\rm Li}_3(z)+2 \ln z \, {\rm Li}_3(z) 
+2 \ln z \, {\rm Li}_3(1-z)
\nonumber \\ && \quad \quad \left. \left.
-2 \ln\left(\frac{1-z}{z}\right) {\rm Li}_3\left(\frac{z-1}{z}\right)
+2 {\rm Li}_4\left(\frac{z-1}{z}\right)
-2 {\rm Li}_4(z) \right] \frac{dz}{1-z} \right\}
\eeqa
and
\beqa
B(\theta)&=&\frac{e^{\theta}}{e^{2 \theta}-1}
\left\{-2 \zeta_2 \zeta_3+2 \zeta_3 \theta^2 
+\left(\frac{3}{2} \zeta_4 -\frac{\theta^4}{6} \right)
\ln\left(e^{\theta}-1\right) \right.
\nonumber \\ && \quad \quad 
+\left(-\frac{3}{2} \zeta_4-2\zeta_3 \theta +\frac{\theta^4}{6} \right)
\ln\left(e^{\theta}+1\right) 
+2 \zeta_3 \left[{\rm Li}_2\left(-e^{-\theta}\right)
+{\rm Li}_2\left(1-e^{-\theta}\right)\right]
\nonumber \\ && \quad \quad
+\frac{2}{3} \theta^3 \left[{\rm Li}_2\left(e^{-\theta}\right)
-{\rm Li}_2\left(-e^{-\theta}\right)\right]
+2\theta^2 \left[{\rm Li}_3\left(e^{-\theta}\right)
-{\rm Li}_3\left(-e^{-\theta}\right)\right]
\nonumber \\ &&  \quad \quad 
{}+4 \theta \left[{\rm Li}_4\left(e^{-\theta}\right)
-{\rm Li}_4\left(-e^{-\theta}\right)\right]
+4 {\rm Li}_5\left(e^{-\theta}\right)
-4 {\rm Li}_5\left(-e^{-\theta}\right)
\nonumber \\ && \left.
+\int_0^{e^{-\theta}}\left[2 \ln^2 z \, {\rm Li}_2(z^2)
-4 \ln z \, {\rm Li}_3(z^2)+3 {\rm Li}_4(z^2)\right] \frac{dz}{1-z^2}
\right\} \, .
\eeqa

As discussed in \cite{GHKM}, and as we have verified, the massless limit of the cusp anomalous dimension, 
i.e. the limit $\theta \rightarrow \infty$, can be written as
\beq
\lim_{\theta \rightarrow \infty} \Gamma_{\rm cusp}=\theta \sum_{n=1}^{\infty}  
\left(\frac{\alpha_s}{\pi}\right)^n K^{(n)}  
\label{massless}
\eeq
where at one loop $K^{(1)}=C_F$,
at two loops $K^{(2)}=C_F K/2$, and at three loops 
\beqa
K^{(3)}&=&C_F C_A^2 \left(\frac{245}{96}-\frac{67}{36} \zeta_2
+\frac{11}{24}\zeta_3+\frac{11}{8}\zeta_4\right)
+C_F C_A T_F n_f \left(-\frac{209}{216}+\frac{5}{9}\zeta_2
-\frac{7}{6}\zeta_3\right)
\nonumber \\ &&
+C_F^2 T_F n_f \left(\zeta_3-\frac{55}{48}\right)
-\frac{1}{27} C_F T_F^2 n_f^2 \, .
\eeqa
We note that the numbers appearing in $K^{(3)}$ can be easily seen explicitly in the expression for $\Gamma^{(3)}$, via Eqs. (\ref{3loop}) and (\ref{c1}), while their origin is less transparent when $\Gamma^{(3)}$  is expressed in terms of harmonic polylogarithms.

We observe that the combination $(K^{(3)}/C_F) \Gamma^{(1)}$ gives the full 
$n_f^2$ term and the full $C_F^2 T_F n_f$ term, as well as parts of the 
$C_F C_A T_F n_f$ and of the $C_F C_A^2 c_1$ terms in Eq. (\ref{3loop}). 
Furthermore, we observe that the remaining $C_F C_A T_F n_f$ terms as well as some further $C_F C_A^2 c_1$ terms can be absorbed into the combination 
$\Gamma^{(2)}-(K/2) \Gamma^{(1)}$ multiplied by an overall $K$.
Thus we can rewrite Eq. (\ref{3loop}), after these observations and some work, 
in the even simpler form 
\beq
\Gamma^{(3)}= C^{(3)} +K^{'(3)} \Gamma^{(1)}
+K \left[\Gamma^{(2)}-\frac{K}{2}\Gamma^{(1)}\right]
\label{gamma3v2}
\eeq
where $K^{'(3)}=K^{(3)}/C_F$ and 
\beqa
C^{(3)}&=& C_F C_A^2 \left\{
\frac{\zeta_2}{2}-\frac{\zeta_3}{8}
-\frac{9}{8}\zeta_4
-\frac{\zeta_2}{2}\theta-\frac{\theta^2}{4}
-\frac{\theta^3}{12}-\frac{\theta^4}{24} \right.
\nonumber \\ && \quad \quad
+\frac{\theta^2}{4} \ln \left(1-e^{-2 \theta}\right)
-\frac{3}{4} \theta {\rm Li}_2\left(e^{-2 \theta}\right)
-\frac{5}{8} {\rm Li}_3\left(e^{-2 \theta}\right)
\nonumber \\ && \hspace{-10mm}
+\coth\theta \left[-\frac{\zeta_3}{4}+\frac{15}{8}\zeta_4
+\left(\frac{\zeta_2}{2}-\frac{\zeta_3}{2}
+\frac{9}{8}\zeta_4\right) \theta 
+\left(\frac{1}{4}+\zeta_2\right)\theta^2
+\left(\frac{1}{12}+\frac{\zeta_2}{3}\right)\theta^3
+\frac{7}{24}\theta^4+\frac{\theta^5}{24} \right.
\nonumber \\ && \quad \quad
+\frac{1}{2} \left(\theta^2+\theta^3\right) \ln \left(1-e^{-2 \theta}\right)
-\frac{3}{4} \theta^2 {\rm Li}_2\left(e^{-2 \theta}\right)
+\frac{1}{4} {\rm Li}_2\left(1-e^{-2 \theta}\right)
\nonumber \\ && \quad \quad \left.
+\frac{1}{4} (1-7\theta) {\rm Li}_3\left(e^{-2 \theta}\right)
+\frac{1}{2} {\rm Li}_3\left(1-e^{-2 \theta}\right)
-\frac{15}{8} {\rm Li}_4\left(e^{-2 \theta}\right) \right]
\nonumber \\ && \hspace{-10mm}
+\coth^2\theta \left[-\frac{\zeta_2 \zeta_3}{2}-\frac{19}{8}\zeta_4+\frac{3}{2}\zeta_5
+\left(\frac{3}{2} \zeta_3-\frac{15}{8} \zeta_4 \right) 
\theta+\left(\frac{\zeta_3}{4}-\zeta_2\right) \theta^2
\right.
\nonumber \\ && \quad \quad
-\frac{2}{3} \zeta_2 \theta^3
-\frac{\theta^4}{4}-\frac{11}{120}\theta^5 
-\left(-\zeta_3+\zeta_2 \theta+\zeta_2 \theta^2
+\frac{\theta^3}{3}+\frac{\theta^4}{6}\right) \ln\left(1-e^{-2 \theta}\right)
\nonumber \\ && \quad \quad
-\theta^2 \ln^2\left(1-e^{-2 \theta}\right)
-\theta \ln^3\left(1-e^{-2 \theta}\right)
-\frac{1}{8} \ln^4\left(1-e^{-2 \theta}\right)
\nonumber \\ && \quad \quad
+\left[\frac{\zeta_2}{2}+\zeta_2\theta-2\theta^2
+\frac{\theta^3}{12}-\theta \ln\left(1-e^{-2 \theta}\right) \right]
{\rm Li}_2\left(e^{-2 \theta}\right)
\nonumber \\ && \quad \quad
-\frac{1}{4} {\rm Li}_2^2\left(e^{-2 \theta}\right)
+\frac{1}{2} \ln^2\left(1-e^{-2 \theta}\right) {\rm Li}_2\left(1-e^{-2 \theta}\right)
+\frac{1}{4} {\rm Li}_2^2\left(1-e^{-2 \theta}\right)
\nonumber \\ && \quad \quad
-\frac{1}{2} \ln^2\left(e^{2 \theta}-1\right)
{\rm Li}_2\left(\frac{1}{1-e^{2 \theta}}\right) 
+\left[\frac{\zeta_2}{2}-\frac{3}{2} \theta -\frac{\theta^2}{4}
-\ln\left(1-e^{-2 \theta}\right)\right]{\rm Li}_3\left(e^{-2 \theta}\right)
\nonumber \\ && \quad \quad
-\left[\theta+\ln\left(1-e^{-2 \theta}\right) \right] 
{\rm Li}_3\left(1-e^{-2 \theta}\right)
-\left[2 \theta+\ln \left(1-e^{-2 \theta}\right) \right] 
{\rm Li}_3\left(\frac{1}{1-e^{2 \theta}}\right)
\nonumber \\ && \quad \quad \left.
-\frac{9}{8} \theta {\rm Li}_4\left(e^{-2 \theta}\right)
+{\rm Li}_4\left(1-e^{-2 \theta}\right)-{\rm Li}_4\left(\frac{1}{1-e^{2 \theta}}
\right)-\frac{3}{2} {\rm Li}_5\left(e^{-2 \theta}\right) \right]
\nonumber \\ && \hspace{-10mm} \left.
+ \frac{1}{4} \left[A(\theta)-A(0)+B(\theta)-B(0)\right] \right\} \, .
\label{C3}
\eeqa
We note that $C^{(3)}$ contains simpler fractions than $c_1$, and it has a 
somewhat shorter expression. In fact the structure appearing 
in Eq. (\ref{gamma3v2}) appears naturally, as the complicated fractions 
in  $c_1$ are absorbed in $K^{(3)}$ and in the combination of lower-loop 
cusp anomalous dimensions. Thus, $C^{(3)}$ is simpler, it is independent 
of $n_f$, and it involves an overall color factor $C_F C_A^2$.

Eq. (\ref{gamma3v2}) is our main and simplest 
expression for the three-loop cusp anomalous dimension.

\mysection{Numerical results and approximations for top-quark production}

We continue with numerical results for the QCD cusp anomalous dimension at one, two, and three loops. Setting the number of colors $N_c=3$, all the color factors can be calculated explicitly. Furthermore, for numerical evaluations one also has to make a choice for the number of light flavors, $n_f$. Since we are mostly interested in top-quark production, we choose $n_f=5$ in all numerical results in this section.

\begin{figure}
\begin{center}
\includegraphics[width=9cm]{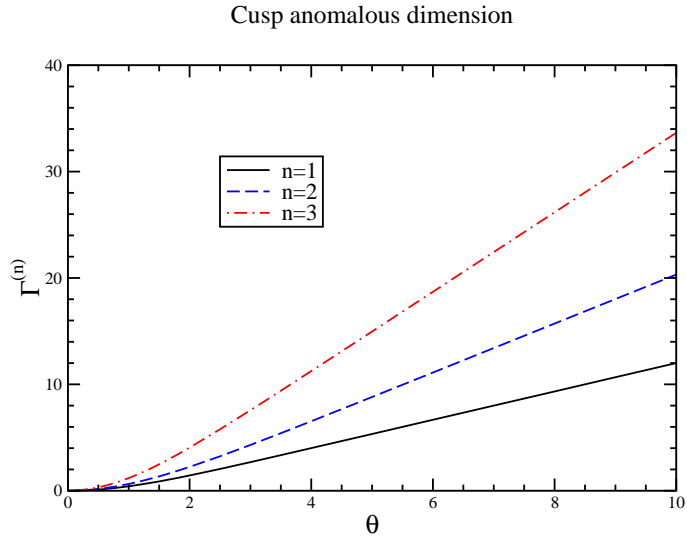}
\caption{The cusp anomalous dimension $\Gamma^{(n)}$ at one, two, and three loops with $n_f=5$ as a function of cusp angle $\theta$.}
\label{gammathetaplot}
\end{center}
\end{figure}

In Figure \ref{gammathetaplot} we plot the cusp anomalous dimension $\Gamma^{(n)}$ at one, two, and three loops as a function of $\theta$, with $n_f=5$. The numerical values increase with $\theta$ for all three curves. Furthermore, for each $\theta$ the values increase with loop order. As expected from Eq. (\ref{massless}), the results become nearly linear at large values of $\theta$.

For small $\theta$ we can  expand the cusp anomalous dimension around $\theta=0$ (see also \cite{NK2loop} and \cite{GHKM}) and we find 
\beq
\Gamma^{(1)}_{\rm exp}=\frac{C_F}{3} \theta^2
\eeq
\beq
\Gamma^{(2)}_{\rm exp}=\left[C_F C_A \left(\frac{47}{54}-\frac{\zeta_2}{3}\right)
-\frac{5}{27} C_F T_F n_f \right] \theta^2
\eeq
\beqa
\Gamma^{(3)}_{\rm exp}&=&\left[C_F C_A^2 \left(\frac{473}{288}-\frac{85}{54}\zeta_2+\frac{5}{72}\zeta_3+\frac{5}{4} \zeta_4\right)
+C_F C_A T_F n_f \left(-\frac{389}{648}+\frac{10}{27} \zeta_2-\frac{7}{18} \zeta_3 \right) \right.
\nonumber \\ && \left.
+C_F^2 T_F n_f \left(-\frac{55}{144}+\frac{\zeta_3}{3} \right)
-\frac{1}{81} C_F T_F^2 n_f^2 \right] \theta^2
\eeqa
where the subscript ``exp'' stands for ``expansion,'' and we neglect higher powers of $\theta$ beyond $\theta^2$. It is interesting to note that $\Gamma^{(n)}$ scales as $\theta^2$ for small $\theta$ as shown in the above equations, while it scales as $\theta$ for large $\theta$ as we saw in Eq. (\ref{massless}). 

We note that in processes involving heavy-quark pair production, it is convenient to express the cusp angle in terms of the quantity $\beta=\sqrt{1-\frac{4m^2}{s}}$ (which at lowest-order is the heavy-quark speed), where $m$ is the heavy-quark mass and $s=(p_i+p_j)^2$ with $p_i^{\mu}$, $p_j^{\mu}$ the quark momenta, via the expression $\theta=\ln[(1+\beta)/(1-\beta)]$. Equivalently we have 
$\beta=\tanh(\theta/2)$. 

For example, the one-loop result, Eq. (\ref{1loop}), can be expressed in terms of $\beta$ as \cite{NK2loop} 
\beq
\Gamma^{(1)}(\beta)=C_F \left[-\frac{(1+\beta^2)}{2 \beta} \ln\left(\frac{1-\beta}{1+\beta}\right)-1\right] \, .
\label{1loopbeta}
\eeq  
Similarly, the two-loop result can be found explicitly as a function of $\beta$ in Ref. \cite{NK2loop}. 

\begin{figure}
\begin{center}
\includegraphics[width=9cm]{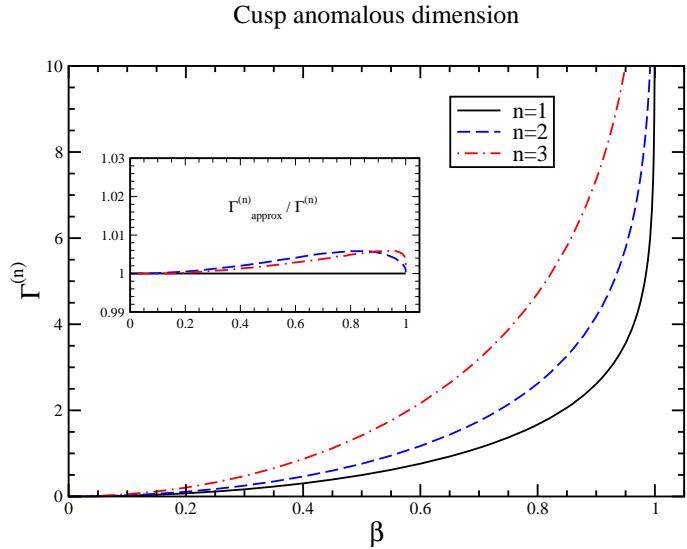}
\caption{The cusp anomalous dimension $\Gamma^{(n)}$ at one, two, and three loops with $n_f=5$ as a function of $\beta$.}
\label{gammabetaplot}
\end{center}
\end{figure}

We note that for small $\theta$, we have $\theta^2=4\beta^2+{\cal O}(\theta^4)$, so the small $\theta$ expansion formulas can easily be rewritten in terms of $\beta$. On the other hand, the massless limit, i.e. the infinite $\theta$ limit, corresponds to $\beta=1$.
The cusp anomalous dimension can also be plotted as a function of $\beta$. In Fig. \ref{gammabetaplot} we plot $\Gamma^{(n)}$ as a function of $\beta$ using $n_f=5$. The curves rise sharply as $\beta \rightarrow 1$.

As first shown in Ref. \cite{NK2loop} for the two-loop case, we can construct approximations valid for all values of $\beta$.
The expansion around $\beta=0$ gives very good 
approximations to $\Gamma^{(n)}$ at small $\beta$. 
The expression in Eq. (\ref{massless}) gives the large $\beta$ limit, which 
shows that in that limit the higher-loop results are proportional to the 
one-loop result. Thus, we can derive an approximation to $\Gamma^{(n)}$ for all 
$\beta$ values by starting with the small $\beta$ 
expansion of $\Gamma^{(n)}$,  
then adding $K^{'(n)} \, \Gamma^{(1)}$ and subtracting from it its small 
$\beta$ expansion:
\beq
\Gamma^{(n)}_{\rm approx}=\Gamma^{(n)}_{\rm exp}
+K^{'(n)} \Gamma^{(1)}-K^{'(n)} \Gamma^{(1)}_{\rm exp}
\label{approx}
\eeq
where $K^{'(n)}=K^{(n)}/C_F$ (and thus $K^{'(1)}=1$, $K^{'(2)}=K/2$).

For the one-loop case, $n=1$, the approximate and exact results are identical. 
Applying Eq. (\ref{approx}) to the higher-loop cases $n=2$ and $n=3$, and numerically evaluating the constants and setting $n_f=5$, we find the very simple expressions
\beq
\Gamma^{(2)}_{\rm approx}(\beta)=
-0.38649 \, \beta^2+1.72704 \; \Gamma^{(1)}(\beta)
\label{Gamma2approx}
\eeq
\beq
\Gamma^{(3)}_{\rm approx}(\beta)=
0.09221 \, \beta^2+2.80322 \; \Gamma^{(1)}(\beta)
\label{Gamma3approx}
\eeq
where $\Gamma^{(1)}(\beta)$ is given by Eq. (\ref{1loopbeta}).

In the inset plot of Fig. \ref{gammabetaplot} we plot the ratio $\Gamma^{(n)}_{\rm approx} / \Gamma^{(n)}$ for $n=1$, 2, 3, as a function of $\beta$ using $n_f=5$. For $n=1$ the ratio is identically 1, as noted above. We see that this very simple approximation works remarkably well for $n=2$ and $n=3$, with the ratio not differing by more than a few per mille from unity for the entire $\beta$ range, and in fact indistinguishable from 1 from much of the $\beta$ range.

It is important to note that the cusp anomalous dimension is the soft anomalous dimension for the process $e^+ e^- \rightarrow t{\bar t}$ \cite{NK2loop}, and it is also the first element of the anomalous dimension matrix that appears in resummations and approximate higher-order calculations \cite{NKtt,N3LO} for top hadroproduction via the processes $q{\bar q} \rightarrow t{\bar t}$ and $gg \rightarrow t{\bar t}$. Therefore its numerical value is important in calculations of cross sections. As the plots show, the numerical value of $\Gamma^{(n)}$ increases with order. However, the overall contribution to $\Gamma_{\rm cusp}$ is moderated by the overall factor $(\alpha_s/\pi)^n$.

\begin{figure}
\begin{center}
\includegraphics[width=9cm]{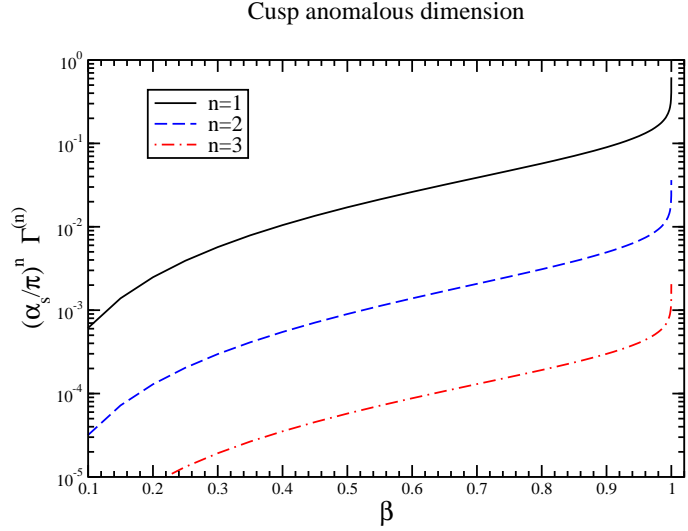}
\caption{The cusp anomalous dimension $(\alpha_s/\pi)^n \, \Gamma^{(n)}$ at one, two, and three loops with $n_f=5$ and $\alpha_s=0.108$ as a function of $\beta$, for $\beta$ values up to  0.999999.}
\label{gammabetaalphaspiplot}
\end{center}
\end{figure}

To illustrate the last point, in Fig. \ref{gammabetaalphaspiplot} we plot the quantities $(\alpha_s/\pi)^n \, \Gamma^{(n)}$ with $n=1$, 2, 3, using $n_f=5$ and $\alpha_s=0.108$. We note that numerically the $t {\bar t}$ cross sections at the LHC and the Tevatron receive most contributions from the region $0.3 < \beta <0.8$. We also observe that, including the $(\alpha_s/\pi)^n$ factors, the one-loop result is about twenty times larger than the two-loop result, and the two-loop result is about fifteen times larger than the three-loop result, as can be seen from Fig. \ref{gammabetaalphaspiplot}.

\mysection{A conjecture for the $n$-loop cusp anomalous dimension}

From our expressions in Section 2, we begin to see a pattern emerging for the cusp anomalous dimension. We write the cusp anomalous dimension at $n$th order in terms of the previous orders plus an extra term $C^{(n)}$. Thus we have
\beq
\Gamma^{(1)}=C^{(1)}
\label{gamma1v3}
\eeq
\beq
\Gamma^{(2)}=C^{(2)}+K^{'(2)} C^{(1)}
\label{gamma2v3}
\eeq
and
\beq
\Gamma^{(3)}=C^{(3)}+2 K^{'(2)} C^{(2)}+K^{'(3)} C^{(1)}
\label{gamma3v3}
\eeq
Note that at one loop $C^{(1)}$ is defined to be $\Gamma^{(1)}$ since we start with $n=1$. At two loops $C^{(2)}$ is simply found from  Eq. (2.3), i.e. it is $(C_F C_A/2)$ times all the terms in curly brackets in Eq. (2.3) [note that $(K/2) \Gamma^{(1)}=K^{'(2)} C^{(1)}$]. At three loops, Eq. (\ref{gamma3v3}) is a rewriting of Eq. (\ref{gamma3v2}) using Eq. (\ref{gamma2v3}). See also the corresponding observation in Ref. \cite{GHKM}.

Observing Eqs. (\ref{gamma1v3}), (\ref{gamma2v3}), (\ref{gamma3v3}), 
we conjecture that the $n$-loop result is given by 
\beq
\Gamma^{(n)}=\sum_{k=1}^n \frac{(n-1)!}{(k-1)! \, (n-k)!} \, K^{'(k)} \, C^{(n-k+1)} \, .
\label{NKconj}
\eeq
It is easy to check that the above expession reproduces the results for $n=1,2,3$. We conjecture that this relation will hold for arbitrary $n$. For example, for $n=4$ we predict
\beq
\Gamma^{(4)}= C^{(4)}+3  K^{'(2)} C^{(3)}+3 K^{'(3)} C^{(2)}+K^{'(4)} C^{(1)}
\eeq
and for $n=5$ we predict
\beq
\Gamma^{(5)}= C^{(5)}+4 K^{'(2)} C^{(4)}+6  K^{'(3)} C^{(3)}+4 K^{'(4)} C^{(2)}
+K^{'(5)} C^{(1)} \, .
\eeq
Note that $C^{(n)}$ has overall color factor $C_F C_A^{n-1}$, and it does not involve $n_f$ terms as first noted in \cite{GHKM}. We predict that the terms in $\Gamma^{(n)}$ will naturally fall into place in the patterns suggested by the conjecture. The conjecture provides the structure of the $n$-loop result, and if true it can be of further use because the calculation of $K^{(n)}$ can predate that of $\Gamma^{(n)}$, and thus $\Gamma^{(n)}$ can be known up to the term $C^{(n)}$ without further calculation. We also note that starting at four loops new non-planar terms arise in the calculation. The non-planar $n_f$ terms would still appear in $K^{(n)}$. It will be interesting to see how such new corrections will satisfy the relations we have found and proposed. 

The above conjecture gives explicit results for $\Gamma^{(n)}$ in terms of the $C^{(k)}$ terms and the massless-limit constants $K^{'(k)}$, but it is nice to have the result directly in terms of the lower-order cusp anomalous dimensions. Using our previous expressions to express the $C$'s in terms of the $\Gamma$'s and 
$K^{'}$'s, and rewriting the equations, we thus find the following explicit expressions for $\Gamma^{(n)}$ through $n=5$:
\beqa
\Gamma^{(2)}&=&C^{(2)}+K^{'(2)} \Gamma^{(1)}
\nonumber \\
\Gamma^{(3)}&=&C^{(3)}+2 K^{'(2)} \Gamma^{(2)}
+\left[K^{'(3)}-2 \left(K^{'(2)}\right)^2\right] \Gamma^{(1)}
\nonumber \\
\Gamma^{(4)}&=&C^{(4)}+3  K^{'(2)} \Gamma^{(3)}
+3\left[K^{'(3)}-2 \left(K^{'(2)}\right)^2\right] \Gamma^{(2)}
\nonumber \\ &&
+\left[K^{'(4)}-6 K^{'(3)}K^{'(2)}+6\left(K^{'(2)}\right)^3\right] \Gamma^{(1)}
\nonumber \\
\Gamma^{(5)}&=&C^{(5)}+4  K^{'(2)} \Gamma^{(4)}
+6\left[K^{'(3)}-2 \left(K^{'(2)}\right)^2\right] \Gamma^{(3)}
\nonumber \\ &&
+4\left[K^{'(4)}-6 K^{'(3)} K^{'(2)}+6\left(K^{'(2)}\right)^3\right] \Gamma^{(2)}
\nonumber \\ &&
+\left[K^{'(5)}-8 K^{'(4)}K^{'(2)}-6\left(K^{'(3)}\right)^2 
+36 K^{'(3)} \left(K^{'(2)}\right)^2 -24 \left(K^{'(2)}\right)^4\right] \Gamma^{(1)}
\label{gamma2345}
\eeqa

We see another pattern emerging, involving repetition of terms in brackets in Eq. (\ref{gamma2345}), when writing the $n$-loop cusp anomalous dimension directly in terms of the lower-loop ones. We can thus rewrite our conjecture as
\beq
\Gamma^{(n)}=C^{(n)}+\sum_{k=2}^n \frac{(n-1)!}{(k-1)! \, (n-k)!} \, F^{(k)} \, 
\Gamma^{(n-k+1)}
\label{NKconjalt}
\eeq
where
\beqa
F^{(2)}&=&K^{'(2)}
\nonumber \\
F^{(3)}&=&K^{'(3)}-2 \left(K^{'(2)}\right)^2
\nonumber \\
F^{(4)}&=&K^{'(4)}-6 K^{'(3)} K^{'(2)}+6\left(K^{'(2)}\right)^3
\nonumber \\
F^{(5)}&=&K^{'(5)}-8 K^{'(4)}K^{'(2)}-6\left(K^{'(3)}\right)^2
+36 K^{'(3)} \left(K^{'(2)}\right)^2 -24 \left(K^{'(2)}\right)^4
\eeqa
etc.

Eqs. (\ref{NKconj}) and (\ref{NKconjalt}) provide two alternate but equivalent forms for our conjecture for the $n$-loop cusp anomalous dimension.

\mysection{Conclusions}

I have provided an explicit and compact expression for the three-loop cusp anomalous dimension in QCD, which is Eq. (\ref{gamma3v2}). The result is based on the calculation of \cite{GHKM}, and is expressed in terms of elementary functions, zeta constants, and ordinary polylogarithms. The massless limit was taken and it agrees with known results. 

Numerical results and approximate expressions were also derived for the cusp anomalous dimension through three loops with $n_f=5$, relevant for top-quark production. The approximate formulas, Eqs. (\ref{Gamma2approx}) and (\ref{Gamma3approx}), are very simple, yet they provide excellent approximations to the exact results at the per mille level or better.

The new expressions provide new insights into the structure of the terms. Relations were derived among the cusp anomalous dimensions through three loops. Observing those relations and an emerging pattern, a conjecture was made for the $n$-loop result in terms of the lower-loop expressions and constants from the massless limit. The conjecture displays the analytical structure of the results and may deepen understanding, it can be used to make predictions for higher-loop results, and it is expressed in two alternate forms, Eqs. (\ref{NKconj}) and (\ref{NKconjalt}). 

The results for the cusp anomalous dimension presented in this paper may be useful in calculations of soft-anomalous dimensions for heavy quark production, such as top-antitop pair and single-top production, and other hard-scattering processes. The very simple approximate formulas derived in this paper also enable very fast numerical calculations with excellent accuracy.

\mysection*{Acknowledgements}
This material is based upon work supported by the National Science Foundation 
under Grant No. PHY 1519606.

\appendix
\mysection{Appendix}

The zeta constants used in our equations are given by  $\zeta_k=\sum_{n=1}^{\infty} 1/n^k$. In particular, $\zeta_2=\pi^2/6$, $\zeta_3=1.202056903\cdots$, $\zeta_4=\pi^4/90$, and $\zeta_5=1.036927755\cdots$. 

The polylogarithm is given by ${\rm Li}_k(z)=\sum_{n=1}^{\infty} z^n/n^k$. 
Note that ${\rm Li}_k(1)=\zeta_k$.

The harmonic polylogarithms \cite{RV} 
of weight $n$ are defined iteratively by
\beq
H_{a_1,a_2,\cdots, a_n}(z)=\int_0^z f_{a_1}(z') \, H_{a_2,\cdots,a_n}(z') \, dz'
\eeq
where
\beq
f_0(z)=\frac{1}{z} \quad \quad
f_1(z)=\frac{1}{1-z} \quad \quad
f_{-1}(z)=\frac{1}{1+z}
\eeq
and the weight 1 harmonic polylogarithms are
\beqa
H_0(z)&=&\ln z
\nonumber \\
H_1(z)&=&\int_0^z \frac{dz'}{1-z'}=-\ln(1-z)
\nonumber \\
H_{-1}(z)&=&\int_0^z \frac{dz'}{1+z'}=\ln(1+z)
\eeqa
For example
\beq
H_{0,1}(z)=\int_0^z f_0(z') \, H_1(z') \, dz'=\int_0^z \frac{1}{z'} \,
(-\ln(1-z'))\, dz'={\rm Li}_2(z) \, .
\eeq

We begin by calculating all the harmonic polylogarithms appearing in the results of Ref. \cite{GHKM} for the cusp anomalous dimension through three loops that involve the variable $y=1-e^{-2\theta}$. The calculations involve up to quintiple integrals for the weight 5 harmonic polylogarithms. We first show explicit results for all harmonic polylogarithms that involve multiple integrals that in the end have explicit expressions in terms of standard functions:
\beqa
H_1(y)&=&2\theta 
\\
H_{1,1}(y)&=&2\theta^2 
\\
H_{0,1}(y)&=&{\rm Li}_2\left(1-e^{-2\theta}\right)
\\
H_{1,1,1}(y)&=&\frac{4}{3}\theta^3
\\
H_{1,0,1}(y)&=&-2\zeta_3+2\zeta_2\theta+2\theta{\rm Li}_2\left(e^{-2\theta}\right)+2{\rm Li}_3\left(e^{-2\theta}\right)
\\
H_{0,1,1}(y)&=&\zeta_3+2\theta^2\ln\left(1-e^{-2\theta}\right)-2\theta{\rm Li}_2\left(e^{-2\theta}\right)-{\rm Li}_3\left(e^{-2\theta}\right)
\\
H_{0,0,1}(y)&=&{\rm Li}_3\left(1-e^{-2\theta}\right)
\\
H_{1,1,1,1}(y)&=&\frac{2}{3}\theta^4
\\
H_{1,1,0,1}(y)&=&3\zeta_4-4 \zeta_3 \theta+2\zeta_2 \theta^2-2\theta {\rm Li}_3\left(e^{-2\theta}\right)-3{\rm Li}_4\left(e^{-2\theta}\right)
\\
H_{0,1,1,1}(y)&=&\zeta_4+\frac{4}{3}\theta^3\ln\left(1-e^{-2\theta}\right)-2\theta^2 {\rm Li}_2\left(e^{-2\theta}\right)-2\theta {\rm Li}_3\left(e^{-2\theta}\right)-{\rm Li}_4\left(e^{-2\theta}\right)
\\
H_{1,0,0,1}(y)&=&-\frac{1}{2}{\rm Li}_2^2\left(1-e^{-2\theta}\right)+2\theta{\rm Li}_3\left(1-e^{-2\theta}\right)
\\
H_{1,0,1,1}(y)&=&-3\zeta_4+2 \zeta_3 \theta+2\theta^2{\rm Li}_2\left(e^{-2\theta}\right)+4\theta{\rm Li}_3\left(e^{-2\theta}\right)+3{\rm Li}_4\left(e^{-2\theta}\right)
\\
H_{0,1,0,1}(y)&=&\frac{11}{4}\zeta_4+(-2\zeta_3+2 \zeta_2 \theta) \ln\left(1-e^{-2\theta}\right)
+\left(-\zeta_2+2\theta^2\right) \ln^2\left(1-e^{-2\theta}\right)
\nonumber \\ &&
{}+\frac{1}{4} \ln^4\left(1-e^{-2\theta}\right) +\left[-\zeta_2+4\theta^2+2\theta 
\ln\left(1-e^{-2\theta}\right)+\ln^2\left(1-e^{-2\theta}\right)\right]{\rm Li}_2
\left(e^{-2\theta}\right)
\nonumber \\ &&
{}+\frac{1}{2}{\rm Li}_2^2\left(e^{-2\theta}\right)+\ln^2\left(e^{2\theta}-1\right) 
{\rm Li}_2\left(\frac{1}{1-e^{2\theta}}\right)
+\left[4\theta+2\ln\left(1-e^{-2\theta}\right)\right] {\rm Li}_3 \left(e^{-2\theta}\right)
\nonumber \\ &&
{}+2\ln\left(1-e^{-2\theta}\right){\rm Li}_3 \left(1-e^{-2\theta}\right)
+\left[4 \theta+2\ln\left(1-e^{-2\theta}\right)\right] {\rm Li}_3\left(\frac{1}{1-e^{2\theta}}\right)
\nonumber \\ &&
{}+2 {\rm Li}_4\left(e^{-2\theta}\right)-2 {\rm Li}_4\left(1-e^{-2\theta}\right)+2{\rm Li}_4\left(\frac{1}{1-e^{2\theta}}\right) 
\\
H_{1,1,1,1,1}(y)&=&\frac{4}{15}\theta^5
\\
H_{0,1,1,1,1}(y)&=&\zeta_5+\frac{2}{3}\theta^4\ln\left(1-e^{-2\theta}\right)-\frac{4}{3}\theta^3 {\rm Li}_2\left(e^{-2\theta}\right)-2\theta^2{\rm Li}_3\left(e^{-2\theta}\right)-2\theta{\rm Li}_4\left(e^{-2\theta}\right)
\nonumber \\ &&
-{\rm Li}_5\left(e^{-2\theta}\right)
\\
H_{1,0,1,1,1}(y)&=&-4\zeta_5+2\zeta_4\theta+\frac{4}{3}\theta^3{\rm Li}_2\left(e^{-2\theta}\right)+4\theta^2 {\rm Li}_3\left(e^{-2\theta}\right)+6\theta {\rm Li}_4\left(e^{-2\theta}\right)
\nonumber \\ &&
+4{\rm Li}_5\left(e^{-2\theta}\right)
\\
H_{1,1,0,1,1}(y)&=&6\zeta_5-6\zeta_4 \theta+2 \zeta_3 \theta^2-2\theta^2 {\rm Li}_3\left(e^{-2\theta}\right)-6\theta {\rm Li}_4\left(e^{-2\theta}\right)-6 {\rm Li}_5\left(e^{-2\theta}\right)
\\
H_{1,1,1,0,1}(y)&=&-4 \zeta_5+6 \zeta_4 \theta -4 \zeta_3 \theta^2
+\frac{4}{3} \zeta_2 \theta^3 +2 \theta {\rm Li}_4\left(e^{-2\theta}\right)
+4 {\rm Li}_5\left(e^{-2\theta}\right)
\eeqa

We continue by showing results for two harmonic polylogarithms of weight 5 involving $y=1-e^{-2\theta}$ where the integrations can be reduced down to a single integral (plus standard functions in the result for the second one):
\beq
H_{1,1,0,0,1}(y)=\int_0^y\left[-\frac{1}{2}{\rm Li}_2^2(z)
-\ln(1-z){\rm Li}_3(z)\right] \frac{dz}{1-z}
\eeq
\beqa
H_{1,0,1,0,1}(y)&=& 6 \zeta_5+\frac{11}{2} \zeta_4 \theta
+\frac{\theta}{2} \ln^4\left(1-e^{-2\theta}\right)  
+(-2\zeta_3+2\zeta_2 \theta) {\rm Li}_2\left(e^{-2\theta}\right)
\nonumber \\ &&
-\ln^3\left(1-e^{-2\theta}\right)  {\rm Li}_2\left(1-e^{-2\theta}\right)
+(2 \zeta_2-4 \theta^2) {\rm Li}_3\left(e^{-2\theta}\right)
\nonumber \\ &&
+3 \ln^2\left(1-e^{-2\theta}\right) {\rm Li}_3\left(1-e^{-2\theta}\right) 
-8 \theta {\rm Li}_4\left(e^{-2\theta}\right)
\nonumber \\ &&
-6 \ln\left(1-e^{-2\theta}\right) {\rm Li}_4\left(1-e^{-2\theta}\right)  
-6{\rm Li}_5\left(e^{-2\theta}\right)
+6 {\rm Li}_5\left(1-e^{-2\theta}\right)
\nonumber \\ && \hspace{-5mm}
+\int_0^y \left[-\ln(1-z) \ln^3 z
+\frac{1}{2} \ln^2(1-z) \ln^2 z 
-\ln^2 z \, {\rm Li}_2(z)  \right. 
\nonumber \\ && \quad \quad 
-\ln(1-z) \ln z \, {\rm Li}_2(1-z) 
+\ln^2\left(\frac{1-z}{z}\right) {\rm Li}_2\left(\frac{z-1}{z}\right)
\nonumber \\ && \quad \quad 
+\frac{1}{2}{\rm Li}_2^2(1-z)+2 \ln z \, {\rm Li}_3(z) 
+2 \ln z \, {\rm Li}_3(1-z)
\nonumber \\ && \quad \quad \left. 
-2 \ln\left(\frac{1-z}{z}\right) {\rm Li}_3\left(\frac{z-1}{z}\right)
+2 {\rm Li}_4\left(\frac{z-1}{z}\right)
-2 {\rm Li}_4(z) \right] \frac{dz}{1-z}
\eeqa

Next we provide results for the harmonic polylogarithms up to weight 5 appearing in the results for the three-loop cusp anomalous dimension in Ref. \cite{GHKM} that involve the variable $x=e^{-\theta}$.
We first show explicit results for all harmonic polylogarithms that have explicit expressions in terms of standard functions:
\beqa
H_1(x)&=&-\ln\left(1-e^{-\theta}\right)
\\
H_{-1}(x)&=&\ln\left(1+e^{-\theta}\right)
\\
H_{1,0}(x)&=&\theta \ln\left(1-e^{-\theta}\right)-{\rm Li}_2\left(e^{-\theta}\right)
\\
H_{-1,0}(x)&=&-\theta \ln\left(1+e^{-\theta}\right)+{\rm Li}_2\left(-e^{-\theta}\right)
\\
H_{1,0,0,0,0}(x)&=&-\frac{\theta^4}{24} \ln\left(1-e^{-\theta}\right)
+\frac{\theta^3}{6} {\rm Li}_2\left(e^{-\theta}\right)
+\frac{\theta^2}{2} {\rm Li}_3\left(e^{-\theta}\right) 
\nonumber \\ &&
+\theta{\rm Li}_4\left(e^{-\theta}\right)+{\rm Li}_5\left(e^{-\theta}\right)   
\\
H_{-1,0,0,0,0}(x)&=&\frac{\theta^4}{24} \ln\left(1+e^{-\theta}\right)
-\frac{\theta^3}{6} {\rm Li}_2\left(-e^{-\theta}\right)
-\frac{\theta^2}{2} {\rm Li}_3\left(-e^{-\theta}\right) 
\nonumber \\ &&
-\theta{\rm Li}_4\left(-e^{-\theta}\right)-{\rm Li}_5\left(-e^{-\theta}\right)   
\eeqa

We finish by showing results for harmonic polylogarithms of weight 5 involving $x=e^{-\theta}$ where the integrations can be reduced down to a single integral:
\beq
H_{1,0,1,0,0}(x)=\int_0^x\left[\frac{1}{2} \ln^2 z \, {\rm Li}_2(z) 
-2 \ln z \, {\rm Li}_3(z) +3 {\rm Li}_4(z)\right] \frac{dz}{1-z}
\eeq
\beq
H_{-1,0,1,0,0}(x)=\int_0^x\left[\frac{1}{2} \ln^2 z \, {\rm Li}_2(z) 
-2 \ln z \, {\rm Li}_3(z) +3 {\rm Li}_4(z) \right] \frac{dz}{1+z}
\eeq
\beq
H_{1,0,-1,0,0}(x)=\int_0^x\left[-\frac{1}{2} \ln^2 z \, {\rm Li}_2(-z)
+2 \ln z \, {\rm Li}_3(-z) -3 {\rm Li}_4(-z)\right] \frac{dz}{1-z}
\eeq
\beq
H_{-1,0,-1,0,0}(x)=\int_0^x\left[-\frac{1}{2} \ln^2 z \, {\rm Li}_2(-z)
+2 \ln z \, {\rm Li}_3(-z) -3 {\rm Li}_4(-z)\right] \frac{dz}{1+z}
\eeq


\begin{thebibliography}{99}

\bibitem{AMP}
A.M. Polyakov, Nucl. Phys. B {\bf 164}, 171 (1980).

\bibitem{BNS}
R.A. Brandt, F. Neri, and M. Sato, Phys. Rev. D {\bf 24}, 879 (1981).

\bibitem{IKR}
S.V. Ivanov, G.P. Korchemsky, and A.V. Radyushkin, 
Yad. Fiz. {\bf 44}, 230 (1986) [Sov. J. Nucl. Phys. {\bf 44}, 145 (1986)].  

\bibitem{KR}
G.P. Korchemsky and A.V. Radyushkin, Phys. Lett. B {\bf 171}, 459 (1986); 
Nucl. Phys. B {\bf 283}, 342 (1987); Phys. Lett. B {\bf 279}, 359 (1992).  

\bibitem{NK2loop}
N. Kidonakis, Phys. Rev. Lett. {\bf 102}, 232003 (2009)
[arXiv:0903.2561 [hep-ph]].

\bibitem{NK2lc}
N. Kidonakis, in Proceedings of DPF2009, arXiv:0910.0473 [hep-ph]. 

\bibitem{CHMS}
D. Correa, J. Henn, J. Maldacena, and A. Sever, JHEP 1205 (2012) 098 
[arXiv:1203.1019 [hep-th]].

\bibitem{HH}
J.M. Henn and T. Huber, JHEP 1309 (2013) 147 [arXiv:1304.6418 [hep-th]].

\bibitem{GHKM}
A. Grozin, J.M. Henn, G.P. Korchemsky, and P. Marquard, 
Phys. Rev. Lett. {\bf 114}, 062006 (2015) [arXiv:1409.0023 [hep-ph]]; 
JHEP 1601 (2016) 140 [arXiv:1510.07803 [hep-ph]].

\bibitem{NKGS}
N. Kidonakis and G. Sterman, Phys. Lett. B {\bf 387}, 867 (1996);
Nucl. Phys. {\bf B505}, 321 (1997) [hep-ph/9705234].

\bibitem{KOS}
N. Kidonakis, G. Oderda, and G. Sterman, Nucl. Phys. {\bf B531}, 365 (1998) 
[hep-ph/9803241].

\bibitem{ADS}
S.M. Aybat, L.J. Dixon, and G. Sterman, Phys. Rev. Lett. {\bf 97},
072001 (2006) [hep-ph/0606254]; 
Phys. Rev. D {\bf 74}, 074004 (2006) [hep-ph/0607309].

\bibitem{BN}
T. Becher and M. Neubert, Phys. Rev. Lett. {\bf 102}, 162001 (2009)
[arXiv:0901.0722 [hep-ph]]. 

\bibitem{GM}
E. Gardi and L. Magnea, JHEP 0903 (2009) 079 [arXiv:0901.1091 [hep-ph]]. 

\bibitem{LD}
L.J. Dixon, Phys. Rev. D {\bf 79}, 091501 (2009) [arXiv:0901.3414 [hep-ph]].

\bibitem{MSS}
A. Mitov, G. Sterman, and I. Sung, Phys. Rev. D {\bf 79}, 094015 (2009)
[arXiv:0903.3241 [hep-ph]].

\bibitem{BFS}
M. Beneke, P. Falgari, and C. Schwinn, Nucl. Phys. B {\bf 828}, 69 (2010) 
[arXiv:0907.1443 [hep-ph]].

\bibitem{FNPY}
A. Ferroglia, M. Neubert, B.D. Pecjak, and L.L. Yang, JHEP 0911 (2009) 062
[arXiv:0908.3676 [hep-ph]].

\bibitem{DGM}
L.J. Dixon, E. Gardi, and L. Magnea, JHEP 1002 (2010) 081 
[arXiv:0910.3653 [hep-ph]].

\bibitem{GLSW}
E. Gardi, E. Laenen, G. Stavenga, and C.D. White, JHEP 1011 (2010) 155
[arXiv:1008.0098 [hep-ph]].

\bibitem{NKst}
N. Kidonakis, Phys. Rev. D {\bf 81}, 054028 (2010) [arXiv:1001.5034 [hep-ph]];
Phys. Rev. D {\bf 82}, 054018 (2010) [arXiv:1005.4451 [hep-ph]]; 
Phys. Rev. D {\bf 83}, 091503(R) (2011) [arXiv:1103.2792 [hep-ph]].

\bibitem{NKtt}
N. Kidonakis, Phys. Rev. D {\bf 82}, 114030 (2010)
[arXiv:1009.4935 [hep-ph]].

\bibitem{GSW}
E. Gardi, J.M. Smillie, and C.D. White, JHEP 1109 (2011) 114 
[arXiv:1108.1357 [hep-ph]]; 
JHEP 1306 (2013) 088 [arXiv:1304.7040 [hep-ph]].

\bibitem{EG}
E. Gardi, JHEP 1404 (2014) 044 [arXiv:1310.5268 [hep-ph]]. 

\bibitem{FGHMW}
G. Falcioni, E. Gardi, M. Harley, L. Magnea, and C.D. White, 
JHEP 1410 (2014) 10 [arXiv:1407.3477 [hep-ph]].  

\bibitem{ADG}
O. Almelid, C. Duhr, and E. Gardi, 
arXiv:1507.00047 [hep-ph].

\bibitem{N3LO}
N. Kidonakis, Phys. Rev. D {\bf 90}, 014006 (2014) [arXiv:1405.7046 [hep-ph]]; 
Phys. Rev. D {\bf 91}, 031501(R) (2015) [arXiv:1411.2633 [hep-ph]].

\bibitem{RV}
E. Remiddi and J.A.M. Vermaseren, Int. J. Mod. Phys. A {\bf 15}, 725 (2000)
[hep-ph/9905237].

\end{thebibliography}
\end{document}